\documentclass[a4paper,11pt]{article}
\pdfoutput=1
\usepackage{jheppub}
\usepackage[T1]{fontenc}
\usepackage[utf8]{inputenc}
\usepackage{physics}
\usepackage{amsmath,amssymb,amsthm,textcomp}
\usepackage{mathrsfs}
\usepackage{bm,bbm,bbold}
\usepackage{mathtools}
\usepackage{enumerate}
\usepackage{hyperref}
\usepackage{subcaption}

\title{States of low energy in the Schwinger effect}

\author{\'Alvaro \'Alvarez-Dom\'inguez,}
\author{Luis J. Garay,}
\author{Mercedes Mart\'in-Benito,}
\author{Rita B. Neves.}

\affiliation{Departamento de F\'isica Te\'orica and IPARCOS, Universidad Complutense de Madrid, Plaza de las Ciencias 1, 28040 Madrid, Spain}

\emailAdd{alvalv04@ucm.es}
\emailAdd{luisj.garay@ucm.es}
\emailAdd{m.martin.benito@ucm.es}
\emailAdd{rneves@ucm.es}

\newcommand{\bfk}{\textbf{k}}
\newcommand{\bfx}{\textbf{x}}
\newcommand{\Wk}{W_{\bfk}}
\newcommand{\Yk}{Y_{\bfk}}

\graphicspath{{./figures/}} 

\abstract{States of low energy in cosmology minimise the energy density when smeared in a chosen time interval. We extend such construction to generic homogeneous (possibly anisotropic) particle creation settings. Focusing on the Schwinger effect, we study the role played by the support of the smearing function and identify the vacua obtained in the limiting cases of small and large time intervals.
We also analyse the spectral properties of the power spectrum and the number of created particles, which are complementary in characterising the vacuum, and investigate the multipolar contributions coming from the anisotropies.}

\preprint{IPARCOS-UCM-23-020}

\begin{document} 
\maketitle
\flushbottom

\section{Introduction} \label{sec:introduction}

Many particle creation phenomena are an effect of the action of a classical, time-dependent external agent which excites a quantum matter field. This is the case, for instance, in cosmological pair production due to an evolving spacetime geometry \cite{Parker1969,Ford1987,Weinberg2008}, and in the Schwinger effect resulting from a strong electric field \cite{Sauter1931,Schwinger1951}. 

In this context, it is commonly assumed that the external agents are strong enough so that backreaction of the quantum test fields can be neglected. Then a mean-field approximation is considered. See \cite{Pla2021,Pla2022} for a recent study on the validity of the semiclassical approach in the Schwinger effect. In this case, for strengths of the electric field below the critical Schwinger limit, which is of the order of $10^{18} \ \text{V/m}$, particle production is exponentially suppressed \cite{Schwinger1951,Yakimenko2018}. In the literature the possibility of reaching such critical strength with intense laser facilities is still under discussion as this would produce fermionic pair cascades which could deplete the electric field \cite{Fedotov2010,Bulanov2010,Gonoskov2013}. Nevertheless, recent works prove that these obstacles might be overcome, making supercritical electric fields attainable in the future \cite{Ilderton2023,Vincenti2019}. For a recent review see \cite{Fedotov2023}. In addition, the Schwinger effect has recently been observed in an analogue mesoscopic experiment in graphene \cite{Schmitt2023}.

Pair production effects are often explained in the context of quantum field theory in curved spacetimes. The Hamiltonians describing these systems depend on time, and thus when canonically quantizing each classical theory one encounters the possibility of constructing infinitely many different quantum theories. Equivalently, there is freedom in the choice of the annihilation and creation operators, and therefore of the quantum vacuum. One might then compare two different selections and find that one particular vacuum is excited with respect to another choice of vacuum: particles have been produced.

Different quantum theories predict different expectation values for physical observables. For example, in standard cosmology there is a natural choice of vacuum for cosmological perturbations, the Bunch-Davies vacuum \cite{Birrell1982}, as it is usually fixed at the onset of inflation, a quasi-de Sitter regime. However, in models with pre-inflationary dynamics, such as Loop Quantum Cosmology (LQC) or other bouncing models, it is well-known that different choices of vacua provide different power spectra (see e.g. \cite{Fahn2018,Elizaga2019,Agullo2015,DeBlas2016,Elizaga2021}). Since this is a quantity that is directly related with observations, it is considered the relevant magnitude in cosmology. On the other hand, works about the Schwinger effect usually study the number of created particles as the magnitude of physical relevance. As a consequence of the ambiguities, in the literature there is still an open discussion about the physical interpretation of the time evolution of the number of created particles \cite{Alvarez2023,Ilderton2022,Yamada2021}. Here, we argue that we need both the power spectrum and the particle number to completely characterise our choice of vacuum. This is why we introduce the notion of power spectrum in the Schwinger effect. In addition, the study of these magnitudes will allow us to learn about the anisotropies introduced by the electric field. Concretely, we find that anisotropies do not play an important role in the power spectrum neither in the infrared nor in the ultraviolet, whereas at intermediate scale multipoles do contribute. The same applies to the particle number.

Many authors have proposed a wide range of choices of vacua depending on the physical properties that they want to imprint on the quantum theory. Adiabatic vacua, first put forward by Parker in \cite{Parker1969} and later formalised by Lüders and Roberts in \cite{Luders1990}, are among the most common. Based on the WKB approximation, they naturally extend to a slowly changing external agent the usual notion of plane waves defining the Minkowski vacuum in flat spacetime. Other options include the diagonalization of the Hamiltonian  \cite{Fulling1979,Fahn2018,Elizaga2019}, instantaneous minimization of the renormalised stress-energy tensor \cite{Agullo2015,Handley2016}, vacua minimizing oscillations of the primordial power spectrum in cosmological scenarios \cite{DeBlas2016,Elizaga2021}, or of the time evolution of the particle number in the Schwinger effect \cite{Dabrowski2014,Dabrowski2016}, unitary implementation of the dynamics \cite{Cortez2015,Cortez2020,Garay2020,Alvarez2021}, among many others.

In this work, we explore the so-called States of Low Energy (SLEs) in the Schwinger effect. Their definition in cosmology was motivated by \cite{Fewster2000}, which showed that the renormalised energy density smeared along a time-like curve is bounded from below as a function of the state. Then, in \cite{Olbermann2007}, this result was applied to generic cosmological models considering smearing functions supported on the wordline of an isotropic observer, and a procedure was devised in order to explicitly obtain such a state. Thus, SLEs are states that minimise the smeared energy density. Here, we investigate the role of the support of the smearing function in the context of the Schwinger effect, and find asymptotic regimes for sufficiently small and large supports, thus providing physical interpretation to the ambiguities in the choice of smearing function.

H. Olbermann proved in \cite{Olbermann2007} one appealing property of SLEs in Friedmann-Lemaître-Robertson-Walker (FLRW) cosmological backgrounds: they are of Hadamard type. This relates to the ultraviolet behaviour of the two-point function, and guarantees that computations such as that of the stress-energy tensor are well defined \cite{Fewster2013}. There are some studies on the Hadamard property in static electric backgrounds \cite{Wrochna2012} and under time-dependent external potentials \cite{Finster2016}. Along these lines, we will see in this work that the ultraviolet behaviour of SLEs in the Schwinger effect is compatible with the Hadamard condition, although this property remains to be rigorously proven. The properties of SLEs for cosmological models were further investigated in \cite{Niedermaier2020}, where they were found to have the same infrared behaviour up to a constant factor for any smearing function and the same ultraviolet behaviour independently of the smearing function. In the context of cosmological perturbations, these authors found that these states are suitable candidates for vacua in models with a period of kinetic dominance prior to inflation, as they provide the correct infrared and ultraviolet behaviours for perturbations at the end of inflation. This prompted the proposal of SLEs as vacua of cosmological perturbations in the context of LQC \cite{Martin-Benito2021-1,Martin-Benito2021-2}. These two works have shown an interesting dependence of SLEs on the smearing function: they are independent of it as long as it is wide enough around the bounce of LQC \cite{Martin-Benito2021-1}, but very sensitive to whether the moment of the bounce is included in the support of the smearing function \cite{Martin-Benito2021-2}. The notion of SLEs has been recently extended to fermionic fields in \cite{Nadal2023}, where they are applied in a radiation-dominated CPT-invariant Universe.

The structure of the paper is as follows. In section \ref{sec:settings} we start by reviewing the issue of quantum vacuum ambiguities, and we extend the construction of SLEs to general homogeneous settings, considering especially the Schwinger effect. In section \ref{sec:sopf} we investigate the dependence on the smearing function. Section \ref{sec:anisotropies} is dedicated to anisotropies. Here, we define the power spectrum in the Schwinger effect and investigate the multipolar contributions. Finally, in section \ref{sec:num} we study the number of created particles for different choices of the smearing function. Section \ref{sec:conclusions} is dedicated to conclusions and closing remarks.

\section{SLEs in general homogeneous settings} \label{sec:settings}

In \cite{Olbermann2007} H. Olbermann introduced a procedure to define and compute the SLEs in FLRW cosmological backgrounds. In this section our aim is to extend this method to generic spatially homogeneous backgrounds, paying special attention to the Schwinger effect. For completeness, let us first summarise how we can parameterise the ambiguities in the choice of quantum vacua.

\subsection{Quantum vacuum ambiguities}

For concreteness, we consider the motion of a classical charged scalar field $\phi(t,\bfx)$ in flat spacetime coupled to an external spatially-homogeneous time-dependent electric field. The dynamics of this matter field is governed by the Klein-Gordon equation
\begin{equation} \label{eq:KG}
    \left[ (\partial_{\mu}+iqA_{\mu})(\partial^{\mu}+iqA^{\mu})+m^2 \right]\phi(t,\textbf{x})=0,
\end{equation}
where $q$ and $m$ are the charge and the mass of $\phi(t,\bfx)$, and $A_{\mu}(t,\bfx)$ is the electromagnetic potential. We use the temporal gauge $A_{\mu}(t)=(0,\textbf{A}(t))$, as this is the only choice which explicitly translates homogeneity to the equations of motion, and thus, to the quantum theory. Each complex Fourier mode
\begin{equation} \label{eq:Fourier}
    \phi_{\bfk}(t)= \frac{1}{(2\pi)^{3/2}}\int d^3\textbf{x} \ e^{-i\bfk\cdot\textbf{x}}\phi(t,\textbf{x})
\end{equation}
satisfies a harmonic oscillator equation 
\begin{equation} \label{eq:harmonic} 
    \ddot{\phi}_{\bfk}(t)+\omega_{\bfk}(t)^2\phi_{\bfk}(t)=0
\end{equation}
with time-dependent frequency
\begin{equation} \label{eq:omegaSchwinger}
    \omega_{\bfk}(t)=\sqrt{|\bfk+q\textbf{A}(t)|^2+m^2}.
\end{equation}
These equations are decoupled for different wavevectors $\bfk$. In addition, note that the real and imaginary parts of each mode ($\phi_{\bfk}^{\text{R}}(t)$ and $\phi_{\bfk}^{\text{I}}(t)$, respectively) also satisfy the same equation, so we can deal with them independently.

Matter fields coupled to other external time-dependent spatially homogeneous agents different from an electric field are also governed by harmonic oscillator equations of the type \eqref{eq:harmonic}. All the information on the external field is encoded in the time-dependent frequency. An example is the case of scalar and tensor gauge invariant perturbations in FLRW backgrounds and in LQC, where now the gravitational field plays the role of the electric field. The following formalism can be applied to these models as well as to particle creation settings which can be described as above. Thus, we will consider a generic frequency, which we denote as $\Omega_{\bfk}(t)$ from now on, reserving $\omega_{\bfk}(t)$ for the particular case of the Schwinger effect \eqref{eq:omegaSchwinger}.

We now wish to canonically quantise the matter field $\phi(t,\bfx)$. First, we choose a basis of solutions $\{\varphi_{\bfk}(t)\}_{\bfk}$ to the dynamic equations \eqref{eq:harmonic}. Note that in general we could consider different basis for $\phi_{\bfk}^{\text{R}}(t)$ and $\phi_{\bfk}^{\text{I}}(t)$, but for simplicity we are fixing the same basis for both of them. In order to preserve the Poisson bracket structure, this basis has to be normalised with respect to the Klein-Gordon product:
\begin{equation} \label{eq:normalization}
    \varphi_{\bfk}(t)\dot{\varphi}^*_{\bfk}(t)-\varphi_{\bfk}^*(t)\dot{\varphi}_{\bfk}(t)=i.
\end{equation}
Then, we can define the quantum field operators $\hat{\phi}^{\alpha}_{\bfk}(t)$ ($\alpha=\text{R},\text{I}$) as the linear combination
\begin{equation}
\label{eq:hatphi}
   \hat{\phi}^{ \alpha}_{\bfk}(t)=\hat{a}^{\alpha}_{\bfk}\varphi_{\bfk}(t)+\hat{a}^{\alpha\dagger}_{\bfk}\varphi_{\bfk}^*(t).
\end{equation}
Different choices of solutions $\varphi_{\bfk}(t)$ translate into different annihilation and creation operators $\hat{a}^{ \alpha}_{\bfk}$ and $\hat{a}^{\alpha\dagger}_{\bfk}$, and vice versa. The quantum vacuum $\ket{0}$ associated with a given choice is defined as the state annihilated by all the annihilation operators; i.e., $\hat{a}^{\alpha}_{\bfk}\ket{0}=0$, for all~$\bfk$. Thus, constructing a particular quantum theory is equivalent to choosing one solution $\varphi_{\bfk}(t)$ to equation~\eqref{eq:harmonic} for each $\bfk$. In other words, all we need to do is to impose initial conditions $(\varphi_{\bfk}(t_0),\dot{\varphi}_{\bfk}(t_0))$ to that equation at a certain time $t_0$. Using the normalisation condition~\eqref{eq:normalization}, we can parameterise the different possible choices (up to an irrelevant phase) as follows~\cite{Mottola2000}:
\begin{equation}
\label{eq:WY}
    \varphi_{\bfk}(t_0)=\frac{1}{\sqrt{2W_{\bfk}(t_0)}}, \qquad \dot{\varphi}_{\bfk}(t_0)=\sqrt{\frac{W_{\bfk}(t_0)}{2}}\left[Y_{\bfk}(t_0)-i\right],
\end{equation}
where $\Wk(t_0)>0$ and $\Yk(t_0)$ are independent real quantities which determine the particular vacuum that we select.

This raises a non-trivial question. If each choice defines a different quantum theory, each leading to different theoretical predictions, which one should we select? Depending on the particular system that we are studying, one can find in the literature many vacuum proposals. For example, in the simplest case in which we have a matter field in flat spacetime where no external agent is present, the frequency $\Omega_{\bfk}$ is constant and the Minkowski vacuum is the only one preserving Poincaré symmetry. This well-known vacuum is defined by a basis $\{\varphi_{\bfk}(t)\}_{\bfk}$ which are positive-frequency plane waves, i.e.,
\begin{equation} \label{eq:planewave}
    \Wk(t_0)=\Omega_{\bfk}, \qquad \Yk(t_0)=0.
\end{equation}
However, when we introduce a time-dependent external agent in the system, it breaks Poincaré invariance and the classical group of symmetries is severely reduced. Then, the criterion of preservation of the classical symmetries in the quantum theory is not strong enough to select one unique vacuum. One possibility is to find the vacuum which minimises the energy density. However, as the Hamiltonian is time-dependent, this criterion has ambiguities. One option is to choose a particular time $t_0$ and find the vacuum that instantaneously minimises the expectation value of the energy density at $t_0$. This prescription defines the so-called instantaneous lowest-energy state (ILES) at $t_0$, given by \cite{Mukhanov2007}
\begin{equation} \label{eq:ILES}
    \Wk(t_0)=\Omega_{\bfk}(t_0), \qquad  \Yk(t_0)=0.
\end{equation}
It is important to remark that the minimum of the energy density only exists if $\Omega_{\bfk}(t_0)^2$ is positive, so we restrict our study to this case.\footnote{Note that in the Schwinger case \eqref{eq:omegaSchwinger} this condition is always verified.} Note that at different times $t_0$ we have different notions of instantaneous lowest-energy states as long as the frequency $\Omega_{\bfk}(t)$ depends on time. In addition, the ILES has another interesting property: it instantaneously diagonalises the Hamiltonian in the basis of the vacuum and its excited states.

\subsection{SLEs in spatially homogeneous backgrounds}\label{sec:SLEshomogeneous}

We introduce in the following the SLEs. With the construction of these states, one aims to minimise the energy density in a finite time interval, instead of at an exact instant of time. More precisely, Fewster \cite{Fewster2000} proved in gravitational scenarios that if we smear the energy density along a timelike curve, it is bounded from below. Then, one can find the particular choice of vacua that provide this minimum: the states of low energy. In \cite{Olbermann2007}, H. Olbermann computed these vacua in FLRW backgrounds and proved that they have the Hadamard property. 

Here we propose a direct generalisation of Olbermann's procedure to systems characterised by modes $\phi_{\bfk}(t)$  satisfying harmonic oscillator equations with time-dependent frequencies $\Omega_{\bfk}(t)$. Let $f(t)$ be a smearing function of compact support $[t_1,t_2]$. Each mode $\phi_{\bfk}(t)$ contributes to the total smeared energy as
\begin{equation} \label{eq:Ek}
    E[\phi_{\bfk}]=\frac{1}{2}\int dt \ f(t)^2\left[ |\dot{\phi}_{\bfk}(t)|^2+\Omega_{\bfk}(t)^2|\phi_{\bfk}(t)|^2 \right].
\end{equation}
The aim is to find, for each $\bfk$, the solution $S_{\bfk}(t)$ which minimises this energy density. The strategy is as follows. First, we provide a fiducial solution $F_{\bfk}(t)$ to the equation of motion. Then, the problem translates into finding complex constants $\lambda_{\bfk}$ and $\mu_{\bfk}$ such that the solution $S_{\bfk}(t)$ is written as the linear combination
\begin{equation} \label{eq:TS}
    S_{\bfk}(t)=\lambda_{\bfk}F_{\bfk}(t)+\mu_{\bfk}F_{\bfk}(t)^*.
\end{equation}
Note that this is actually a Bogoliubov transformation, so in order to preserve the Poisson algebra of the corresponding annihilation and creation operators, the Bogoliubov coefficients should satisfy $|\lambda_{\bfk}|^2-|\mu_{\bfk}|^2=1$. On the other hand, the phase of the solution $S_{\bfk}(t)$ is irrelevant, so without loss of generality we can assume that $\mu_{\bfk}$ is a positive real constant. Substituting \eqref{eq:TS} in the smeared energy density \eqref{eq:Ek}, we can write
\begin{equation} \label{eq:ETk}
    E[S_{\bfk}]=\left( 1+2\mu_{\bfk}^2\right)E[F_{\bfk}]+2\mu_{\bfk}\Re{\lambda_{\bfk}C[F_{\bfk}]},
\end{equation}
where the complex constant $C[F_{\bfk}]$ depends on the fiducial solution $F_{\bfk}(t)$ as
\begin{equation} \label{eq:ck}
    C[F_{\bfk}]=\frac{1}{2}\int dt \ f(t)^2\left[ \dot{F}_{\bfk}(t)^2+\Omega_{\bfk}(t)^2F_{\bfk}(t)^2 \right].
\end{equation}
Direct inspection of \eqref{eq:ETk} reveals that the minimum of $E[S_{\bfk}]$ is reached for the most negative value that the quantity $\Re{\lambda_{\bfk}C[F_{\bfk}]}$ can attain. This is achieved when the principal arguments satisfy $\text{Arg\,}\lambda_{\bfk}+\text{Arg\,}C[F_{\bfk}]=\pi$. Then, using the relation $|\lambda_{\bfk}|^2-|\mu_{\bfk}|^2=1$ we can write $E[F_{\bfk}]$ in \eqref{eq:ETk} only in terms of the Bogoliubov coefficient $\mu_{\bfk}$. Finally, we minimise $E[F_{\bfk}]$ with respect to $\mu_{\bfk}$ and obtain
\begin{equation} 
\label{eq:mulambda} 
\mu_{\bfk}=\sqrt{\frac{E[F_{\bfk}]}{2\sqrt{E[F_{\bfk}]^2-|C[F_{\bfk}]|^2}}-\frac{1}{2}}, \qquad
\lambda_{\bfk}=-e^{-i\arg{C[F_{\bfk}]}}\sqrt{\mu_{\bfk}^2+1}.
\end{equation}
These two coefficients define the SLE~$S_{\bfk}(t)$ through the Bogoliubov transformation \eqref{eq:TS}. 

The construction of the SLE that we presented here seems to be explicitly dependent on the fiducial solution $F_{\bfk}(t)$. However, the SLE is independent of this choice \cite{Niedermaier2020}. On the other hand, the previous deduction does not determine a unique vacuum but a family of SLEs, each one associated with a particular smearing function $f(t)$. In section \ref{sec:sopf} we study this dependence in detail. A natural question is whether there exists a particular SLE which minimises the smeared energy density for all smearing functions $f(t)$. Reference~\cite{Olbermann2007} analysed this question for FLRW spacetimes and proved that such a state exists only when the scale factor is constant, but not otherwise. For a state of minimal energy to exist in a generic background, the frequencies $\Omega_{\bfk}(t)$ of the harmonic oscillator equations must be time-independent. Indeed, according to the Bogoliubov transformation \eqref{eq:TS}, a solution $F_{\bfk}(t)$ is a SLE if and only if the coefficient $C[F_{\bfk}]$ given in \eqref{eq:ck} vanish. Moreover, if we impose that $F_{\bfk}(t)$ is a SLE for all smearing functions, then $\dot{F}_{\bfk}(t)^2+\Omega_{\bfk}(t)^2F_{\bfk}(t)^2=0$, which is compatible with the equation of motion if and only if the frequency is constant. Note that, in the Schwinger effect, the frequency~\eqref{eq:omegaSchwinger} is constant only when the electric field vanishes. In other words, a notion of state of minimal energy does not exist when we apply an electric field.

\subsection{The Schwinger effect}

\begin{figure}[t]
    \centering
    \begin{subfigure}[b]{0.49\textwidth}
    \centering
    \includegraphics[width=\textwidth]{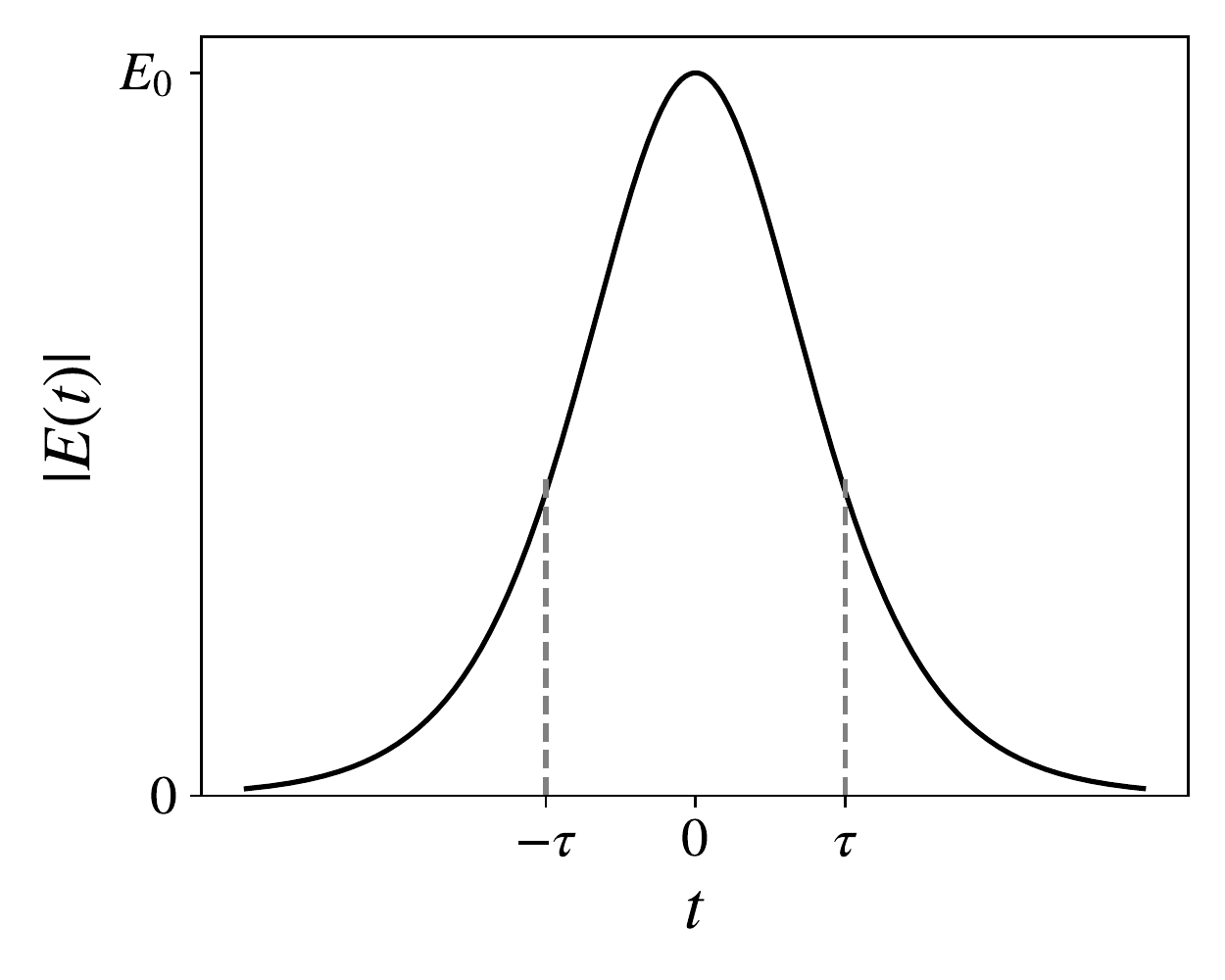}
    \caption{}
    \label{fig:elec}
    \end{subfigure}
    \begin{subfigure}[b]{0.49\textwidth}
    \centering
    \includegraphics[width=\textwidth]{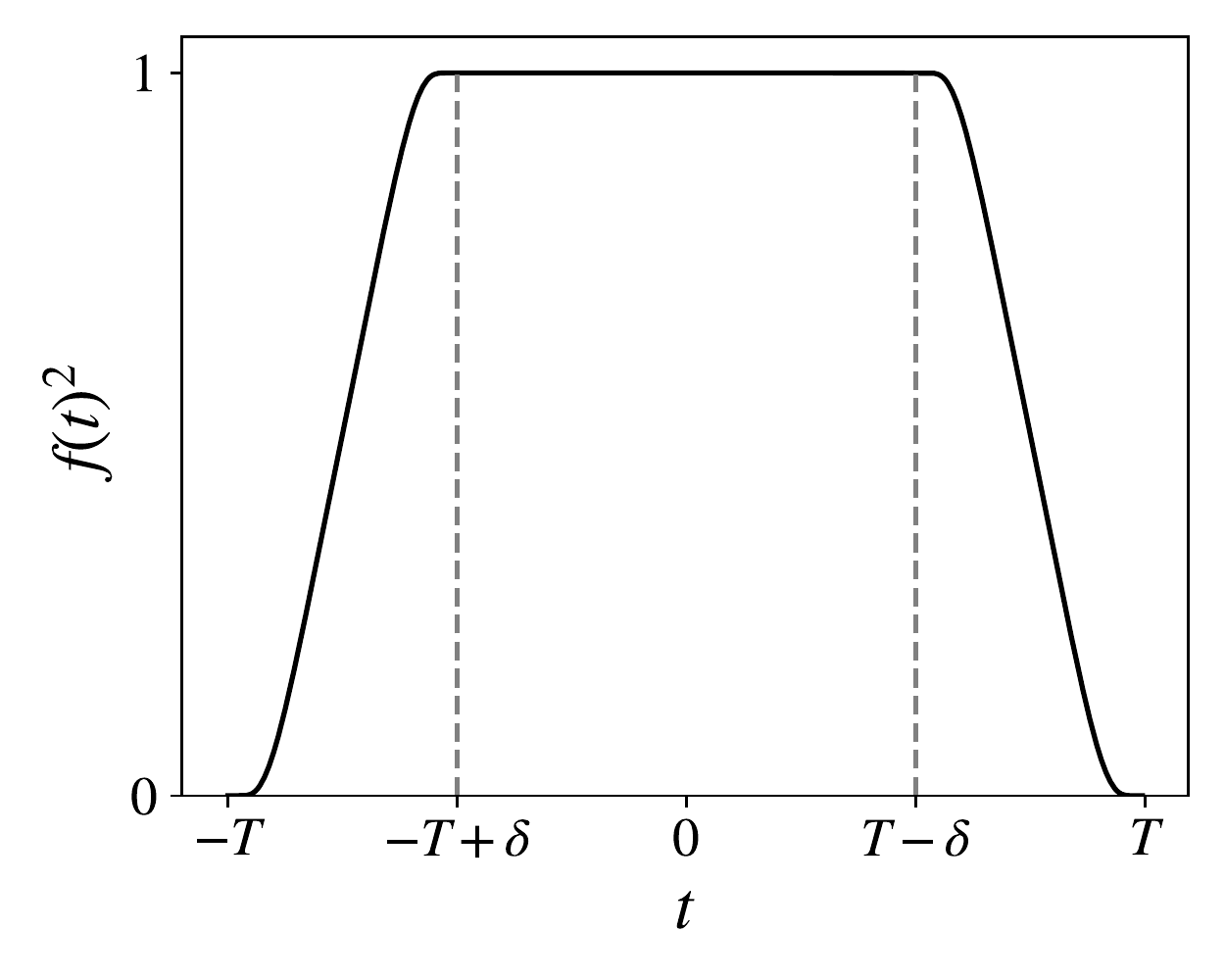}
    \caption{}
    \label{fig:testf}
    \end{subfigure}
    \caption{Illustrative plots of (a) Sauter-type electric field of time width $\tau$ and maximum amplitude $E_0$ and (b) smearing function \eqref{eq:windowf} of compact support $[-T,T]$ and slope of length $\delta$.}
\end{figure}

In the following sections we focus our analysis on the Schwinger effect. In particular, we consider the so-called Sauter-type potential \cite{Sauter1931}, i.e., an electric field potential of the form
\begin{equation}
\label{eq:Sauter}
\textbf{A}(t)=E_0\tau\left[\tanh\left(t/{\tau}\right)+1 \right] \ \textbf{e}_z.
\end{equation}
As it is shown in figure \ref{fig:elec}, it models a P{\"o}schl-Teller electric pulse of maximum amplitude~$E_0$ at time $t=0$ \cite{Poschl1933}. It vanishes asymptotically, and the characteristic width of the pulse is given by~$\tau$. 

This particular potential allows us to find an expression for the in-solution $\varphi_{\bfk}^{\text{in}}(t)$ to equation \eqref{eq:harmonic}, which behaves as a plane wave of frequency $\omega_{\bfk}^{\text{in}}~=~\sqrt{k^2+m^2}$ for $t\rightarrow -\infty$, where $k=|\bfk|$. According to \cite{BeltranPalau2019}, this in-solution can be written in terms of hypergeometric functions \cite{Abramowitz1965}:
\begin{equation}
\label{eq:insolution}
    \varphi_{\bfk}^{\text{in}}(t)=\frac{1}{\sqrt{2\omega_{\bfk}^{\text{in}}}}e^{-i\omega_{\bfk}^{\text{in}}t}\left( 1+e^{2t/\tau} \right)^{(1-i\delta)/2} {}_2F_1\left( \rho_{\bfk}^+,\rho_{\bfk}^-;1-i\tau\omega_{\bfk}^{\text{in}};-e^{-2t/\tau} \right), 
\end{equation}
where 
\begin{equation}
    \delta=\sqrt{(2qE_0\tau^2)^2-1}, \qquad \rho_{\bfk}^{\pm}=\frac{1}{2}\left[ 1-i\tau\left(\omega_{\bfk}^{\text{in}}\pm \omega_{\bfk}^{\text{out}}\right)-i\delta \right], 
\end{equation}
and the out-frequency $\omega_{\bfk}^{\text{out}}=\sqrt{|\bfk+2qE_0\tau \textbf{e}_z|^2+m^2}$ corresponds to the asymptotic limit $t\rightarrow +\infty$ of $\omega_{\bfk}(t)$ in \eqref{eq:omegaSchwinger}. 

To compute the SLEs we could take this in-solution as the fiducial solution $F_{\bfk}(t)$, and thus we could write the SLEs in terms of integrals of hypergeometric functions. However, recall that the construction of the SLEs is independent of the fiducial solution chosen, and thus one may choose any convenient one. For example, for our numerical computations we took the numerical solution with zeroth-order adiabatic initial conditions \cite{Birrell1982} at $t=0$.

In the following sections we use the natural units $c=\hbar=1$, time will be always expressed in units of the time width~$\tau$ of the Sauter-type pulse, and frequencies in units of~$\tau^{-1}$. On the other hand, in our plots, we fix the value of the mass to $m=\tau^{-1}$ and the maximum amplitude of the electric field to $qE_0=\tau^{-2}$. This corresponds to the critical Schwinger limit $qE_0=m^2$ \cite{Yakimenko2018}. For lower strengths, the probability of pair production is exponentially suppressed. Only for electric fields of this order does the Schwinger effect become physically relevant. In practice, the qualitative behaviour of the system for stronger electric fields is the same, and provides no additional information that is relevant to this work. Furthermore, we are interested in studying the physical differences between choices of vacua. Considering larger intensities than the Schwinger limit makes these differences less clear. Finally, it is worth noting that the Schwinger limit is not attainable experimentally yet, although recent works in laser facilities are optimistic about achieving this in the near future \cite{Vincenti2019,Ilderton2023,Ilderton2013}.

\section{Role of the smearing function} \label{sec:sopf}

In the previous section we saw that each SLE minimises the energy density smeared with a certain compact support function $f(t)$. We are interested in studying the physical interpretation of choosing different supports for the smearing function, each defining a particular notion of SLE. As the Sauter potential \eqref{eq:Sauter} is symmetric around its maximum at $t=0$, it will be useful to consider smearing functions with compact support $[-T,T]$, where $T>0$. In particular, we are going to use smooth window functions as shown in figure \ref{fig:testf}. We will describe them in terms of regularised step functions $\Theta_{\delta}(t)$ of width $\delta$, such that in the limit $\delta\to0$ we recover the discontinuous Heaviside step function. The function $\Theta_{\delta}(t)$ interpolates between $0$ and $1$ for $t\in(-\delta/2,\delta/2)$ and it is constant outside. We choose for the interpolating function $\bm{(}1+\tanh\{\cot[\pi(1/2-t/\delta)]\}\bm{)}/2$
although the results will not qualitatively depend on this particular selection. Then, we can write the smearing functions as
\begin{equation} \label{eq:windowf}
    f(t)^2=\left[\Theta_{\delta}(t+T-\delta/2)+\Theta_{\delta}(-t+T-\delta/2)\right]/2.
\end{equation}
We fix a small step width of $\delta=10^{-4}\tau$ for all the figures in this work. For supports smaller than this width (i.e., $T<\delta$), we readapt the parameter by setting $\delta = T/2$ so that it is still smooth.

For simplicity, we are choosing to maintain the shape of the test function, considering only the effects of changing its support. In principle its shape may also be relevant to the resulting SLE. However, for large enough supports the SLEs should be fairly insensitive to the form of the test function, as long as it is reasonably behaved, as is indeed corroborated in \cite{Martin-Benito2021-1}. Furthermore, even when the form of the test function may be relevant, different shapes would simply translate to more or less weight being given to specific time periods when computing the smeared energy density. Therefore, we may understand the physics behind the consequences of different shapes by understanding the physical interpretation of the support first. Besides, one may also argue that more intricate shapes are less natural choices that would require additional motivation.

We saw in the previous section that the freedom in the choice of vacuum is parameterised by $W_{\bfk}(t_0)$ and $Y_{\bfk}(t_0)$, which define via \eqref{eq:WY} the initial conditions of the selected basis of solutions at time $t_0$. We are going to fix $t_0=0$, the instant at which the Sauter-type electric field reaches its maximum, and consider the smearing functions \eqref{eq:windowf}, varying $T$. In addition, in this section we focus on modes whose wavevectors $\bfk$ are parallel to the direction of the electric field. Anisotropies will be analysed in detail in the following sections.

\begin{figure}[t]
    \centering
    \begin{subfigure}{0.49\textwidth}
        \centering
        \includegraphics[width=\textwidth]{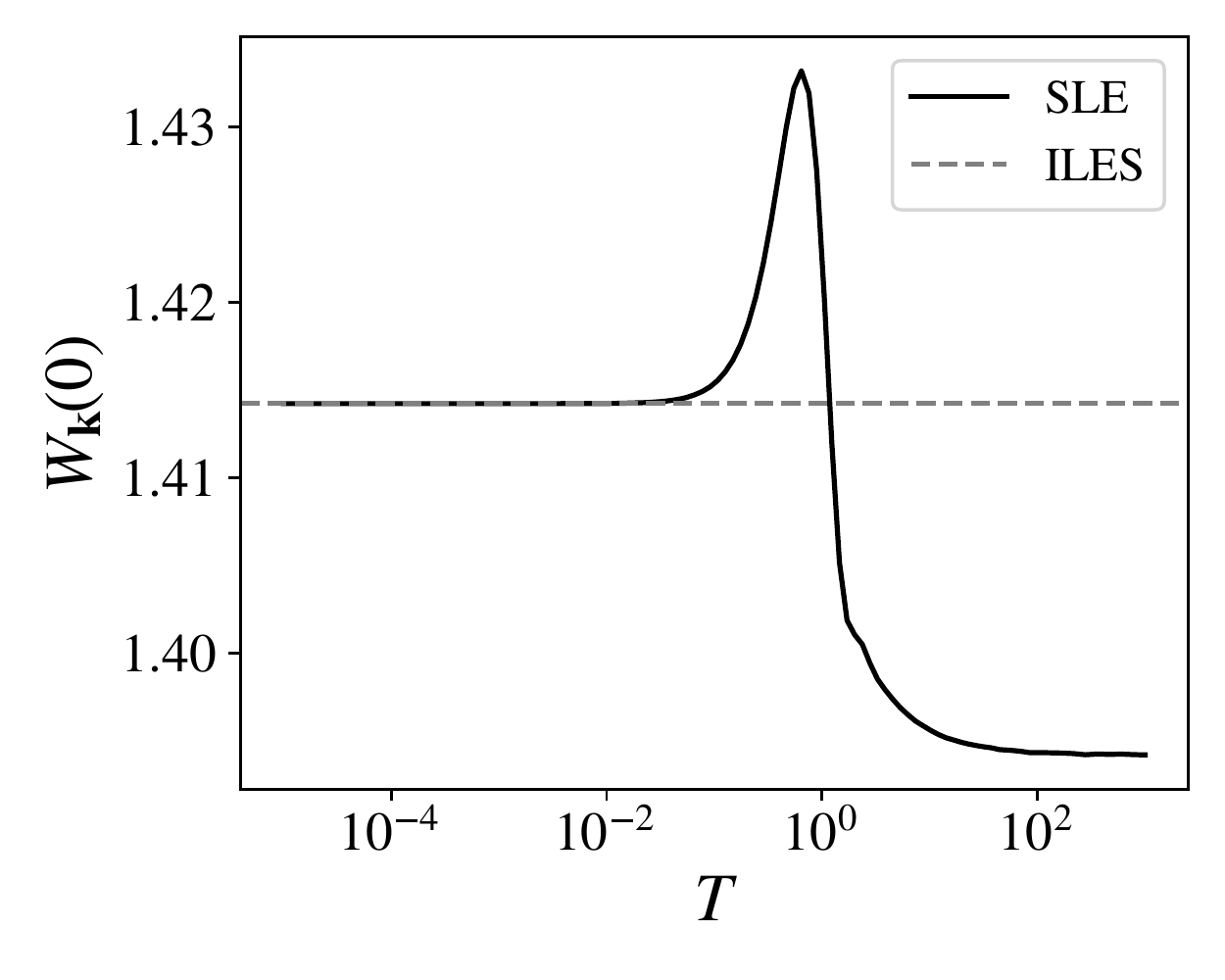}
        \caption{}
        \label{fig:WIR}
    \end{subfigure}
    \begin{subfigure}{0.49\textwidth}
        \centering
        \includegraphics[width=\textwidth]{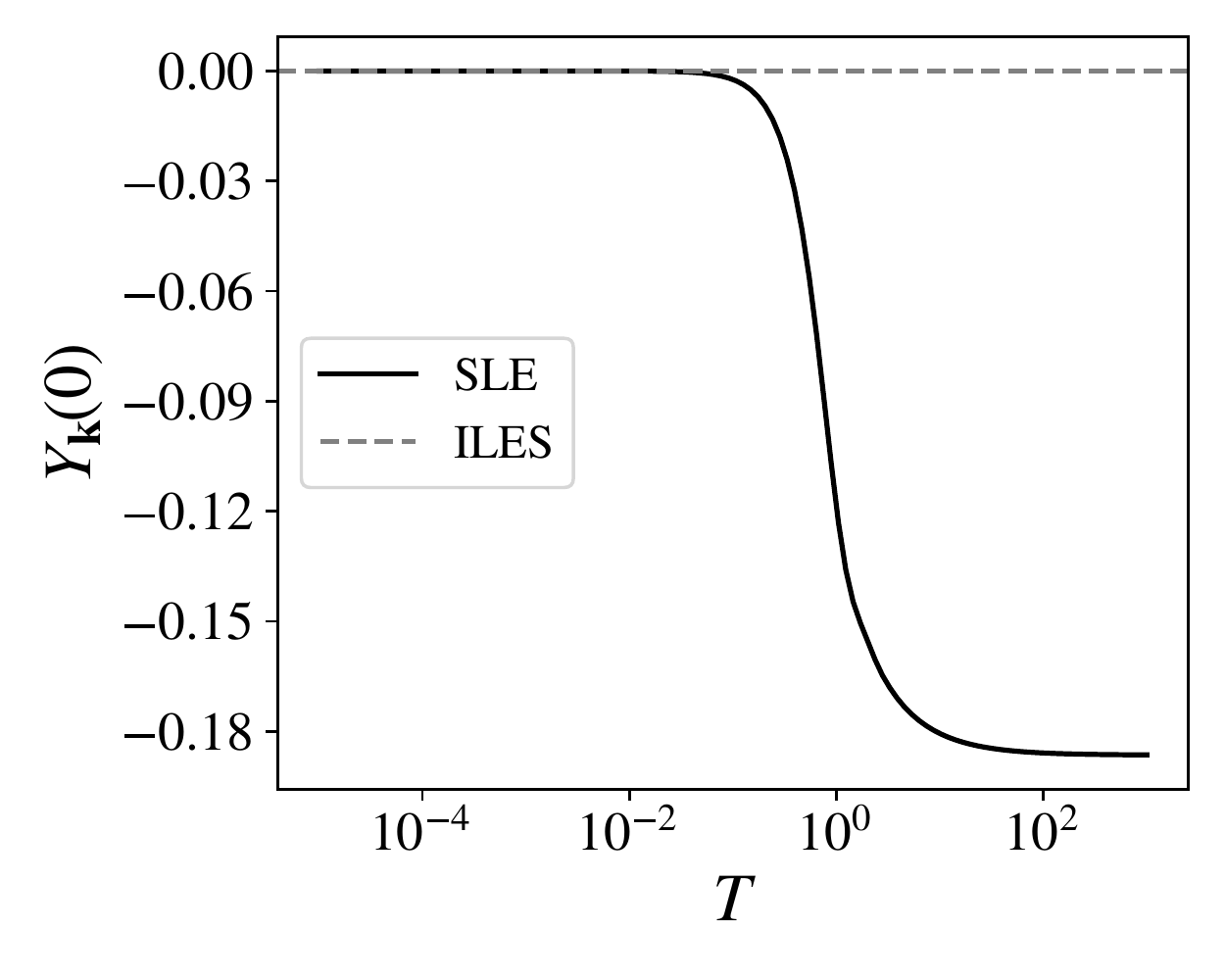}
        \caption{}
        \label{fig:YIR}
    \end{subfigure}
    \caption{Dependence of (a) $W_{\bfk}(t_0=0)$ and (b) $Y_{\bfk}(t_0=0)$ defining the SLEs on the support $[-T,T]$ of the smearing functions \eqref{eq:windowf}. We show the infrared mode whose wavevector~$\bfk$ is parallel to the electric field and $k=10^{-5}\tau^{-1}$. 
 We use units $\tau=1$.}
    \label{fig:WYIR}
\end{figure}

In figure \ref{fig:WYIR} we show $W_{\bfk}(t_0)$ and $Y_{\bfk}(t_0)$ for an infrared mode  with $k=10^{-5}\tau^{-1}$ as functions of the support of the smearing functions. We identify a transition regime around the time scale $\tau$, which is the characteristic length of the Sauter-type electric pulse, where the dependence on the support is not monotonic. It separates the behaviours of the SLEs for small and large supports. We have verified that this happens independently of the strength of the electric field.

When the support is small ($0<T\ll \tau$), the SLEs asymptotically approach the values \eqref{eq:ILES} that characterise the ILES at $t_0$. The physical justification of this fact resides in the definition of the ILES at $t_0$, which minimises the instantaneous energy density obtained identifying the smearing function $f(t)^2$ with the Dirac delta $\delta(t-t_0)$ in \eqref{eq:Ek}. We might then say that the ILES at $t_0$ is the limit for small supports around $t_0$ of the SLEs. However, note that this limit is singular in the sense that the Dirac delta is not a smooth compact-support function, so ILESs are not a particular example of a SLE. These conclusions are also valid for other times different from $t_0$, as we have verified numerically.

On the other hand, we also find an asymptotic constant behaviour for large supports ($T \gg \tau$). This is consistent with the fact that the leading contributions to the smeared energy density are for times in the interval $[-\tau,\tau]$, and that the electric pulse decreases asymptotically. This limit defines a precise vacuum with a well-defined interpretation: the state which minimises the energy density when it is smeared over the entire pulse. 

For other values of $k$ we also distinguish analogue behaviours of $W_{\bfk}(t_0)$ and $Y_{\bfk}(t_0)$ for small and large supports. However, as we increase $k$, the dependence on the support decreases. Indeed, the limit $k\rightarrow \infty$ corresponds to local flat spacetime with no electric field, thus all vacua tend to the Minkowski vacuum defined in \eqref{eq:planewave}. Nevertheless, how fast or slow we reach the Minkowski vacuum strongly depends on each particular vacuum. We will analyse this in more detail in section \ref{sec:num} when studying the number of created particles.

Finally, one might wonder why the noticeable dependence of the SLEs on the support of the test function for supports of the order of the characteristic length of the electric pulse seems absent in the case of LQC \cite{Martin-Benito2021-1}. Indeed, in that work, SLEs are described as independent of the support as long as it is large enough, which agrees with the large support convergent behaviour we observe. Let us then clarify that, although in LQC the equivalent to our potential is different, it also has a characteristic time scale (around the bounce), in which the variations of the potential are most important. This scale plays the same role as our $\tau$, and there it should be less than a hundredth of a Planck second.\footnote{If we approximate the time-dependent mass in the equation of motion of cosmological perturbations in LQC by a P{\"o}schl-Teller potential, the equivalent to $\tau$ is easily found as the time after the bounce at which the potential reduces to half its maximum.} Therefore, the considered supports in \cite{Martin-Benito2021-1} were already quite larger than this scale and the dependence of the SLE on them was minimal, and achieved convergence quickly. In general, the behaviour of SLEs in LQC most likely displays an intermediate regime as we observe in the Schwinger effect, though it corresponds to very small supports around the bounce, which are not physically interesting within the context of cosmology.

\section{Anisotropic power spectrum}\label{sec:anisotropies}

In this section we consider the extension of the common notion of power spectrum in cosmological scenarios to the Schwinger effect. In addition, we are interested in studying in detail the anisotropies present in this electric background. Motivated by the works in anisotropic cosmologies as Bianchi I \cite{Agullo2022}, we introduce an expansion of the power spectrum in Legendre polynomials and analyse its multipolar contributions.

The Wightman function is defined as the vacuum expectation value
\begin{equation}
    \label{eq:Wightman} W(t,\bfx;t',\bfx')=\bra{0}\hat{\phi}^{\dagger}(t,\bfx)\hat{\phi}(t',\bfx')\ket{0}= 2\int \frac{d^3\bfk}{(2\pi)^3} \ e^{i\bfk\cdot (\bfx  -\bfx')} \varphi_{\bfk}(t)\varphi^*_{\bfk}(t'),
\end{equation} 
where we followed the notation of section \ref{sec:settings}. In the last equality we used the definition of the quantum field operator $\hat{\phi}(t,\bfx)$ in terms of the chosen basis $\varphi_{\bfk}(t)$ given by \eqref{eq:hatphi}. Note that the Wightman function only depends on the position vectors through the difference $\bfx-\bfx'$ because the electric field is spatially homogeneous. Writing the integral in \eqref{eq:Wightman} in spherical coordinates, we can integrate out the azimuthal angle. Indeed, we are assuming that the electric field is applied in the $z$ direction and thus it introduces anisotropy only in the polar angle $\theta$. In addition, taking the limit of coincidence $t\rightarrow t'$ yields 
\begin{equation}
    \lim_{t\rightarrow t'} W(t,\bfx;t',\bfx')=\int \frac{dk}{k} \int d(\cos{\theta}) \ e^{i\bfk\cdot (\bfx  - \bfx')} \mathcal{P}(t,\bfk),
\end{equation}
where we defined the power spectrum as
\begin{equation}
\label{eq:PS}
    \mathcal{P}(t,\bfk)=\frac{k^3}{2\pi^2}|\varphi_{\bfk}(t)|^2.
\end{equation}

The power spectrum \eqref{eq:PS} depends on the solutions $\varphi_{\bfk}(t)$ that we choose to construct the quantum theory. More precisely, at a precise time $t_0$ it knows about the ambiguities in the selection of $W_{\bfk}(t_0)$ in \eqref{eq:WY}, although it is oblivious to $Y_{\bfk}(t_0)$. As we are going to see in section \ref{sec:num}, the number of created particles does depend on both $W_{\bfk}(t_0)$ and $Y_{\bfk}(t_0)$. Thus, the power spectrum at a fixed time does not encode all the information about the vacuum. Furthermore, compared with the spectrum of $W_{\bfk}(t_0)$, the infrared power spectrum blurs the differences between different vacua as a consequence of the factor of $k^3$ in its definition~\eqref{eq:PS}.

\begin{figure}
    \centering
    \begin{subfigure}{0.49\textwidth}
        \includegraphics[width=\textwidth]{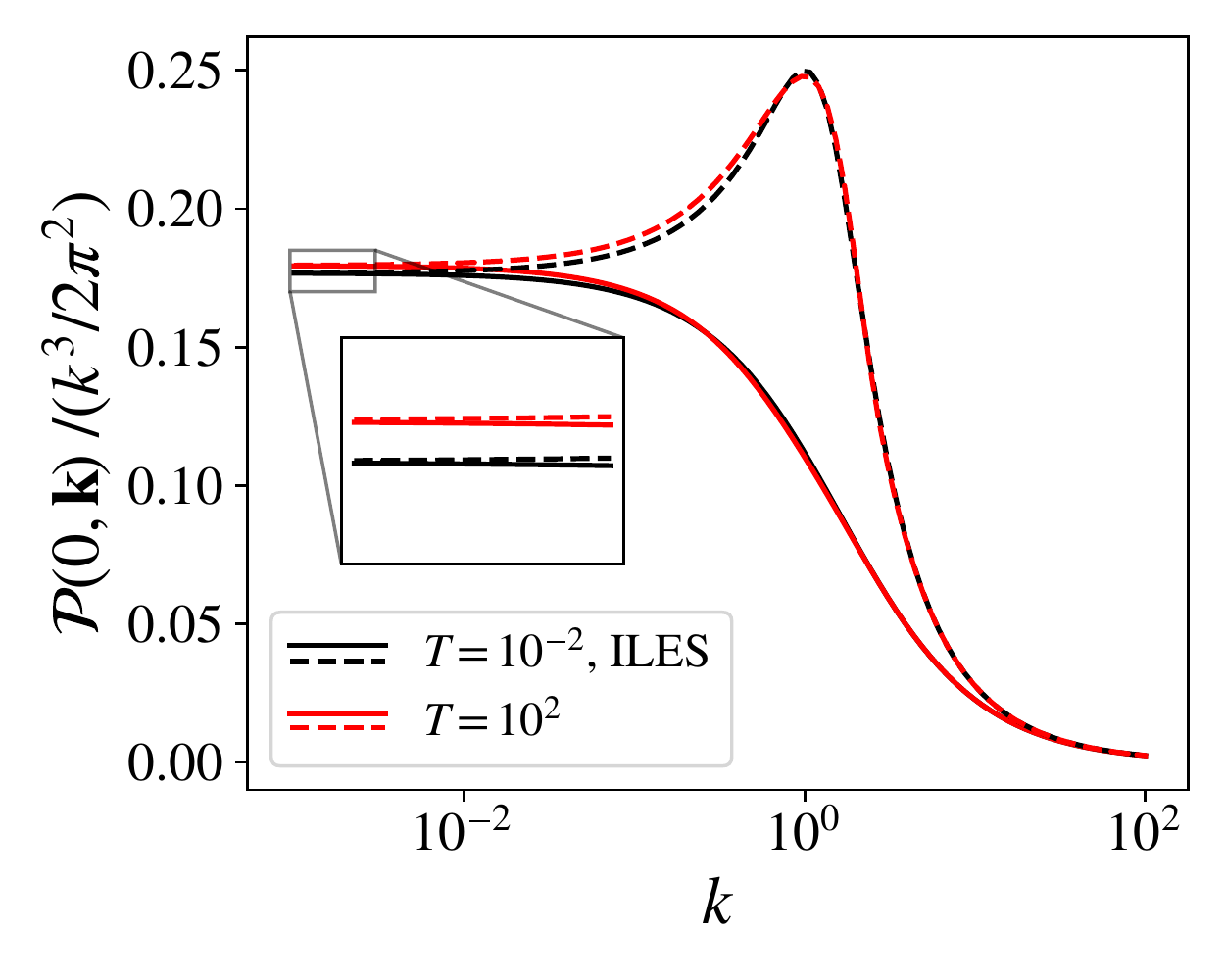}
        \caption{}
        \label{fig:PS}    
    \end{subfigure}
    \begin{subfigure}{0.49\textwidth}
        \includegraphics[width=\textwidth]{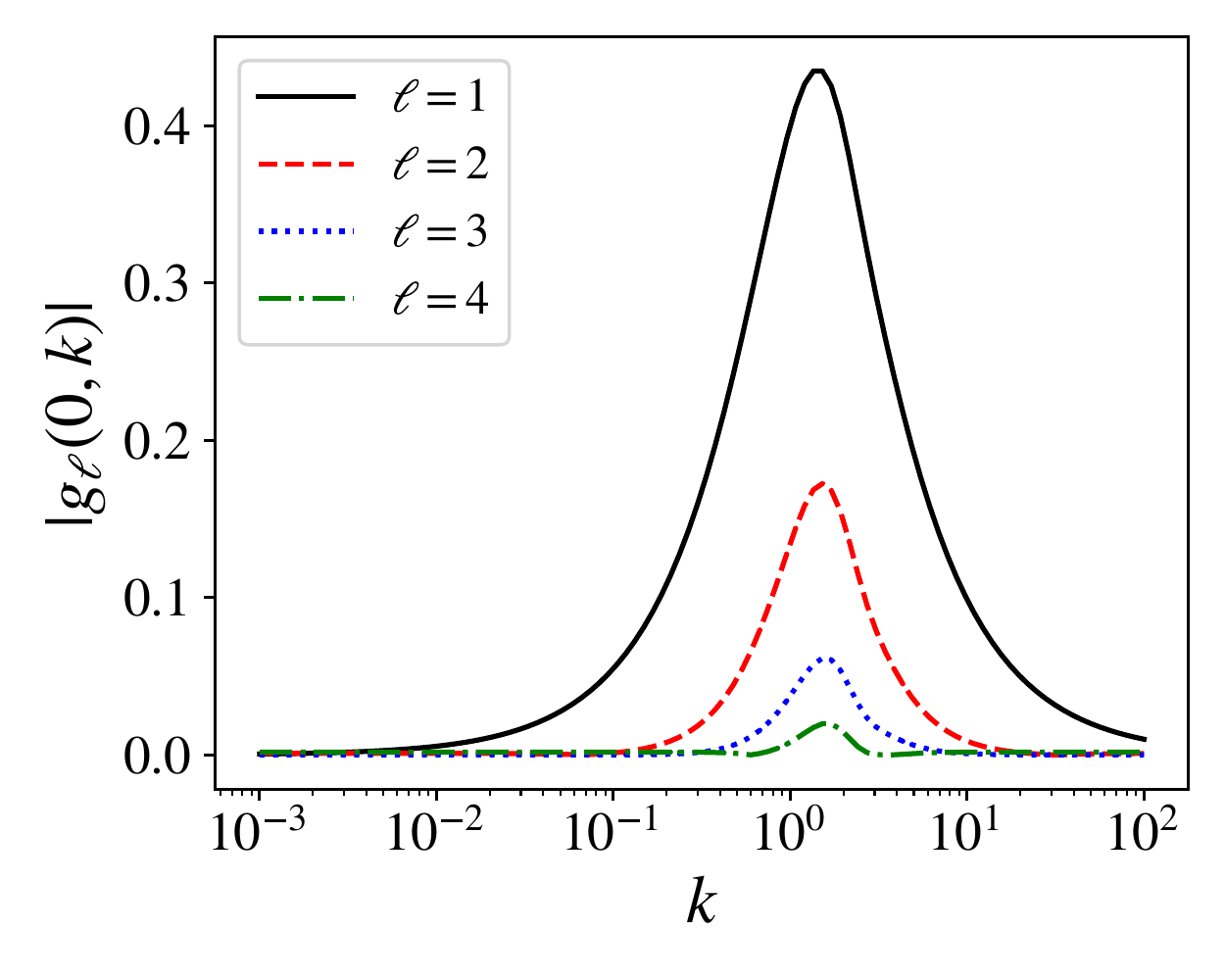}
        \caption{}
        \label{fig:gl}    
    \end{subfigure}
    \caption{(a) Power spectrum divided by $k^3/(2\pi^2)$ at $t= 0$ for a mode parallel (solid) and antiparallel (dashed) to the electric field and for SLEs with different supports. The power spectra for ILES coincide with those for the SLE of smallest support. (b) Absolute value of the contributions $g_{\ell}$ of the multipoles $\ell$ with respect to the monopole at $t = 0$ for a SLE with support $T=10^2\tau$. Note that $g_{\ell}$ are negative for odd values of $\ell$ and positive for even~$\ell$. We use units $\tau=1$.}
    \label{fig:PSgl}
\end{figure}

We show in figure \ref{fig:PS} the power spectrum $\mathcal{P}(t_0,\bfk)$ divided by the factor $k^3/2\pi^2$. This magnitude is computed for SLEs with smearing functions of the type \eqref{eq:windowf} of sufficiently small ($T=10^{-2}\tau$) and sufficiently large ($T=10^2\tau$) supports.\footnote{These are chosen according to figure \ref{fig:WYIR}. This figure refers to a particular infrared mode, but we have verified that the two supports considered here are also sufficiently small and sufficiently large for intermediate and ultraviolet modes as well.} We see that all SLEs have the same infrared behaviour except for a constant. This is in agreement with reference~\cite{Niedermaier2020}. In the ultraviolet, all vacua see a vanishing electric field at sufficiently short scales. Accordingly, they all converge to the same Minkowski vacuum at all times.

Additionally, to investigate the anisotropies we represent modes parallel and antiparallel to the direction of the electric field (i.e., $\theta=0$ and $\theta=\pi$, respectively). Both the infrared and ultraviolet behaviours are oblivious to the direction of $\bfk$. This is rooted in the angular dependence of the frequency in \eqref{eq:omegaSchwinger}, $\omega_\bfk(t)^2 = k^2 + 2q A(t) k \cos \theta + q^2 A(t)^2+ m^2$. Indeed, for $k~\ll~(q^2 A(t)^2~+~m^2)~/~|2 q A(t)|$ and $k \gg 2 |q A(t)|$ the angular contribution is negligible. Conversely, this defines an intermediate regime where the dependence on $\theta$ is important. Accordingly, in figure \ref{fig:PS} the difference between parallel and antiparallel modes is significant at these intermediate scales. Note that in this regime the effects of the anisotropy are much more relevant than that of different choices of SLE. Furthermore, the curves for  $\theta = \pi$ are non-monotonic in contrast with those for $\theta = 0$. Indeed, for positive $\cos \theta$, $\omega_\bfk(t_0)^2$ grows monotonously as $k$ increases,  leading to a power spectrum that monotonously decreases. On the other hand, for negative $\cos \theta$, ${\omega_\bfk}(t_0)^2$ presents a minimum at $k=q A(t_0) |\cos \theta|$, which translates into a maximum in the power spectrum around that point (in our case, $k = \tau^{-1}$). Note that in this work we have chosen $q$ and $A(t)$ to have same sign. Had we chosen them with opposite signs, the roles of $\theta = 0$ and $\theta = \pi$ would have been interchanged. 

We now expand the power spectrum \eqref{eq:PS} in the orthonormal basis of square-integrable functions in $[-1,1]$ formed by the Legendre polynomials, $P_{\ell}(\cos{\theta})$:
\begin{equation}
    \mathcal{P}(t,\bfk)=\sum_{\ell=0}^{\infty} \mathcal{P}_{\ell}(t,k)P_{\ell}(\cos{\theta}),
\end{equation}
where the multipoles are given by
\begin{equation}
    \mathcal{P}_{\ell}(t,k)=\frac{2\ell+1}{2}\int_0^{\pi} d{(\cos{\theta})} \ \mathcal{P}(t,\bfk)P_{\ell}(\cos{\theta}).
\end{equation} 
Let us now consider the multipolar contributions $\ell\geq 1$ with respect to the isotropic monopole $\ell=0$, i.e., the coefficients
\begin{equation}
    g_{\ell}(t,k)=\mathcal{P}_{\ell}(t,k)/\mathcal{P}_0(t,k).
\end{equation}
We show in figure \ref{fig:gl} how these coefficients depend on the module $k$ of the wavevector for the SLE with large support $T=10^2\tau$. We verified that similar behaviours are obtained for smearing functions with different supports. We observe that the maximum contribution of all the multipoles with respect to the monopole happens precisely for the same scale, which is in the aforementioned intermediate regime identified also in figure \ref{fig:PS}. In addition, we confirm that the contribution of multipoles decreases asymptotically in both the infrared and the ultraviolet.

\section{Number of created particles} \label{sec:num}

As mentioned in the previous section, the power spectrum at a fixed time does not fully encapsulate all the information on the vacuum. In cosmology, this is usually the only relevant quantity as it is the only one that can be related with observations of the CMB. However, in general and especially in the context of the Schwinger effect, this can be complemented with the number of created particles in one vacuum with respect to a reference one. 

We will consider as the reference vacuum $\ket{0}^{\textrm{in}}$ the one determined by the in-solution $\varphi_{\bfk}^{\textrm{in}}(t)$ in \eqref{eq:insolution}, which is a positive-frequency plane wave in the asymptotic past. Any other choice of basis of solutions $\varphi_{\bfk}(t)$ defines another quantum theory, with its corresponding annihilation and creation operators $\hat{a}^{\alpha}_{\bfk}$ and $\hat{a}^{\alpha\dagger}_{\bfk}$ according to \eqref{eq:hatphi}. The number of particles in this vacuum with respect to the in-vacuum is
\begin{equation}
\label{eq:Nkak}
    N^{\alpha}_{\bfk}= {}^{\textrm{in}}\!\bra{0}\hat{a}^{ \alpha\dagger}_{\bfk}\hat{a}^{ \alpha}_{\bfk}\ket{0}^{\textrm{in}}=|\beta_{\bfk}|^2,
\end{equation}
where the $\beta$-coefficients quantify the differences between the two solutions:
\begin{equation}
    \beta_{\bfk} = i\left[ \varphi_{\bfk}(t) \dot{\varphi}_{\bfk}^{\textrm{in}}(t) - \dot{\varphi}_{\bfk}(t)\varphi_{\bfk}^{\textrm{in}}(t)\right].
\end{equation}
Noticeably, at each time $t$ this depends on $\varphi_{\bfk}(t)$ as well as its derivative, and therefore encodes information on both $W_{\bfk}(t)$ and $Y_{\bfk}(t)$ of the parameterisation \eqref{eq:WY}. However, it is still not fully descriptive of the vacuum, as it only depends on a combination of these two functions. As such, it may be used in addition to the power spectrum in order to characterise a given vacuum at a given time.

An interesting property of SLEs is that its predicted number of created particles per mode is proportional to the relative difference between the quantum and classical energy. More precisely,
\begin{equation}
    N^{\alpha}_{\bfk}=\frac{1}{2}\frac{{}^{\textrm{in}}\!\bra{0}E[\hat{\phi}^{\alpha}_{\bfk}]\ket{0}^{\textrm{in}}-E[\varphi_{\bfk}]}{E[\varphi_{\bfk}]}.
\end{equation}
This is only true when $\varphi_{\bfk}$ is taken to be a SLE and $\hat{\phi}^{\alpha}_{\bfk}$ is quantized according to \eqref{eq:hatphi}. This property follows from the fact that, as it was commented in section \ref{sec:SLEshomogeneous}, SLEs are the only states such that the constant $C[\varphi_{\bfk}]$ defined in \eqref{eq:ck} vanishes.

\begin{figure}
    \centering
    \includegraphics[width=0.5\textwidth]{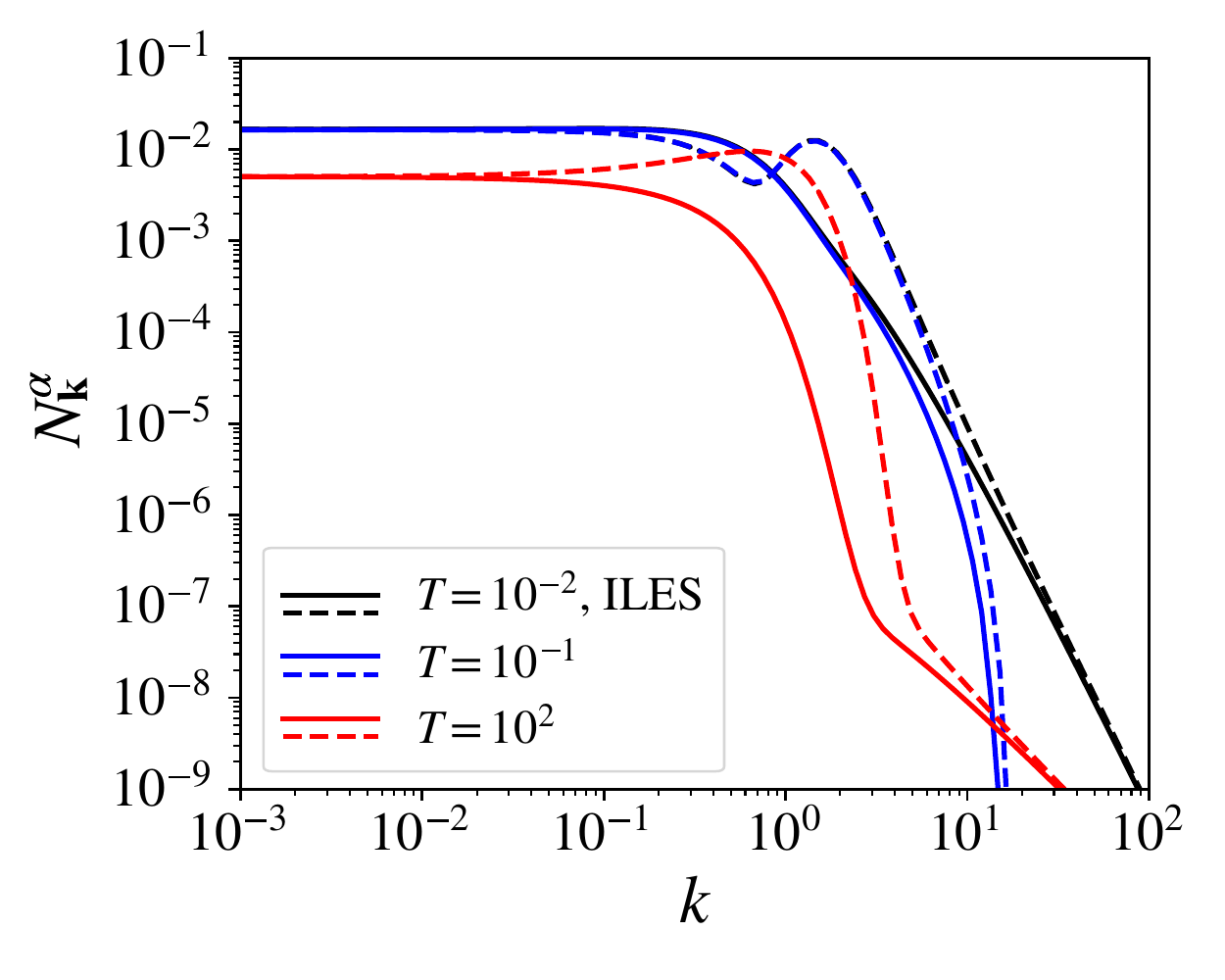}
    \caption{Number of created particles $N^{\alpha}_{\bfk}$ as a function of the module $k$ of the wavevector for SLEs with small and large supports, for modes parallel (solid lines) and antiparallel (dashed lines) to the electric field. For ILES at $t_0$, $N^{\alpha}_{\bfk}$ coincides with that of the SLE with the smallest support considered. We use units $\tau=1$.}
    \label{fig:Nk}
\end{figure}

Figure \ref{fig:Nk} shows the behaviour of the number of created particles $N^{\alpha}_{\bfk}$ for modes parallel and antiparalel to the electric field, as a function of the wavenumber $k$ and for SLEs of sufficiently small ($T = 10^{-2}\tau$) and sufficiently large ($T = 10^{2}\tau$) supports around $t_0$. Again, we identify the same infrared behaviour for all vacua, which are distinguished by a constant contribution. In the ultraviolet, however, each vacuum tends to the Minkowski state at a different rate. For small supports, the spectral particle number $N^{\alpha}_{\bfk}$ of the SLE seems to agree with that of the ILES (see section \ref{sec:sopf}). However, for small enough scales, these states behave differently. To illustrate this separation, we have also represented a SLE with $T=10^{-1}\tau$, whose $N^{\alpha}_{\bfk}$ departs from that of the ILES at a lower (numerically achievable)~$k$. This behaviour is compatible with SLEs being of Hadamard type, while the ILES is not. In fact, Hadamard states are infinite-order adiabatic vacua~\cite{Pirk1993}, whose $N^{\alpha}_{\bfk}$ decays with a power of $k$ proportional to its adiabatic order. Thus, for the ILES the $N^{\alpha}_{\bfk}$ is not exponentially suppressed, decaying more slowly than SLEs for sufficiently ultraviolet modes, not depicted in figure~\ref{fig:Nk}. Along these lines, the~$N^{\alpha}_{\bfk}$ for the SLE with large support  $T = 10^2\tau$ must also decay faster than that for the ILES, for sufficiently ultraviolet modes.\footnote{Solutions to the equation of motion are oscillatory, with increasing frequency after the maximum of the electric pulse, as well as for increasing $k$. Thus, the computation of the SLE becomes computationally demanding for large supports and large $k$, as it requires the integration of oscillations with very short periods.}

Finally, figure \ref{fig:Nk} also shows the intermediate regime where anisotropies are important. As motivated in the previous section, we verify that in the infrared and ultraviolet, the particle number is isotropic. For intermediate scales, parallel modes $\bfk$ to the electric field show a monotonic $N^{\alpha}_{\bfk}$, in contrast to antiparallel modes.

\section{Conclusions} \label{sec:conclusions}

In \cite{Olbermann2007}, SLEs were introduced in general cosmological spacetimes as the states that minimise the energy density, smeared along the trajectory of an isotropic observer. They were shown to be Hadamard states, and later proven to be good candidates for the vacuum of cosmological perturbations in models with a period of kinetic dominance prior to inflation~\cite{Niedermaier2020}. Since then, they have been applied in the context of LQC \cite{Martin-Benito2021-1,Martin-Benito2021-2}, where it was found that they heavily depend on the choice of smearing function only in regards to whether its support includes or excludes the bounce of LQC. Recently, they have also been applied to fermionic fields in a radiation-dominated CPT-invariant universe \cite{Nadal2023}.

In this work, we have extended the construction of SLEs to general spatially homogeneous settings, with the emphasis on the Schwinger effect. To investigate the dependence of these SLEs on the choice of smearing function, we have considered regularised step-like smearing functions with a wide range of supports centred at the maximum of a Sauter-type electric pulse. We discern two asymptotic behaviours of SLEs. In the limit of small supports they behave as ILESs, which instantaneously minimise the energy density (although ILESs are not a particular case of SLEs, just a limiting behaviour). For very large supports the dependence on the support of the smearing function gradually disappears, thus determining in the limit a vacuum which minimises the smeared energy density over the entire electric pulse. For supports of the order of the characteristic time scale of the electric pulse there is a non-trivial dependence. We have been able to draw parallels with what is observed in \cite{Martin-Benito2021-1}, and conclude that the sizes of the support considered in that work already corresponded to the large support regime, which is why convergence is obtained quickly there and no non-trivial dependence on the smearing function is observed.

We have also calculated the power spectrum in the Schwinger effect, analogously to the usual definition in cosmology. We have shown that all SLEs have the same infrared behaviour except for a constant contribution, in agreement with \cite{Niedermaier2020}. In the ultraviolet, all vacua tend to the Minkowski vacuum although at different rates. As the power spectra only depend on the configuration of the state, they all converge for large wavenumbers. However, as the particle number encodes information not only on the configuration of the state but also on its velocity, each vacuum leads to different decay rates when approaching short scales. In particular, we observe that the particle number for all SLEs decays faster than that for the ILES. This might be an indication of SLEs being Hadamard in the Schwinger effect.

Finally, we have analysed the anisotropy of the system. We find that in both the ultraviolet and the infrared regions, the anisotropies do not contribute to either the power spectrum or the number of created particles. An intermediate regime where they are most important has been identified.

\acknowledgments

This work has been supported by Project. No. MICINN PID2020-118159GB-C44 from Spain. R. N. acknowledges financial support from Funda\c{c}\~ao para a Ci\^encia e a Tecnologia (FCT) through the research grant SFRH/BD/143525/2019.

\bibliographystyle{JHEP}
\bibliography{SLEs}

\end{document}